\documentclass[
    ,final            
  ]
  {aipproc}

\layoutstyle{6x9}

\begin{document}

\def\GeV{{\rm GeV}}

\title{A Variable-Flavour-Number Scheme at NNLO}


\author{Robert S. Thorne}{
  address={Cavendish Laboratory, University of Cambridge,
           Madingley Road, Cambridge, CB3 0HE, UK}
}

\begin{abstract}
I present a formulation of a Variable Flavour Number Scheme for heavy quarks
that is implemented up to NNLO in the strong coupling constant and may be 
used in NNLO global fits for parton distributions.  
\end{abstract}

\maketitle


While up, down and strange quarks are treated as effectively  
massless partons, charm, bottom and top 
have to be regarded as heavy partons. 
There are two distinct regimes for these types of quarks. At low scales, 
$Q^2\sim m_H^2$, they are only 
created in the final state and described using the Fixed 
Flavour Number Scheme (FFNS)
$$
F_i(x,Q^2)=C^{FF}_{i,k}(Q^2/m_H^2)\otimes f^{n_f}_k(Q^2).
$$
However, for 
$Q^2 \gg m_H^2$, the coefficient functions contain large 
$\ln(Q^2/m_H^2)$ terms, spoiling the perturbative expansion. 
In this regime it is more appropriate to treat the quarks 
like massless partons,
and the large $\ln(Q^2/m_H^2)$ terms are summed via the  
DGLAP evolution equations. The simplest recipe  
involving this regime is the
Zero Mass Variable Flavour Number Scheme (ZMVFNS).
This ignores all ${\cal O}(m_H^2/Q^2)$ corrections, i.e.
$$
F_i(x,Q^2) = C^{ZMVF}_{i,j}\otimes f^{n_f+1}_j(Q^2).\nonumber
$$

The partons in different flavour-number regions are related  
perturbatively,
$$
f^{n_f+1}_k(Q^2)= A_{jk}(Q^2/m_H^2)
\otimes f^{n_f}_k(Q^2),\nonumber
$$
where the perturbative matrix elements $A_{jk}(Q^2/m_H^2)$
containing $\ln(Q^2/m_H^2)$ terms  
guarantee the correct evolution for both descriptions.
At LO, i.e. zeroth order in $\alpha_S$, the relationship between the two
descriptions is trivial -- $q(g)^{n_f+1}_k(Q^2)\equiv q(g)^{n_f}_k(Q^2).$
At NLO, i.e. first order in $\alpha_S$ ($h^+(Q^2)=(h+\bar h)(Q^2)$),  
$$
h^+(Q^2)= \frac{\alpha_S}{4\pi} P^0_{qg} 
\otimes g^{n_f}(Q^2)\ln\biggl(\frac{Q^2}{m_H^2}\biggr), \quad g^{n_f+1}(Q^2)= 
\biggl(1-\frac{\alpha_S}{6\pi}\ln\biggl(\frac{Q^2}{m_H^2}\biggr)\biggr)
g^{n_f}(Q^2),
$$
i.e. the heavy flavour evolves from zero at $Q^2=m_H^2$ 
and the gluon loses 
corresponding momentum. It is natural to choose $Q^2=m_H^2$ as the 
transition point. At NNLO, i.e. second order in $\alpha_S$,
there is much more complication
$$
f_i^{n_f+1}(Q^2)= \biggl(\frac{\alpha_S}{4\pi}\biggr)^2 \sum_{ij}  
(A^{2,0}_{ij}+A^{2,1}_{ij}\ln(Q^2/m_H^2)+A^{2,2}_{ij}\ln^2(Q^2/m_H^2))
\otimes f_j^{n_f}(Q^2),\nonumber
$$
\noindent where $A^{2,0}_{ij}$ is generally non-zero \cite{buza}. 
There is no longer a smooth transition at this order, and in fact 
the heavy parton begins with a negative value at small $x$. 

This leads to discontinuities in the partons and, without the correct 
treatment, also in the structure functions. 
ZMVFNS coefficient functions also lead 
to discontinuities at the transition point due to a sudden change in the 
flavour number in the coefficient functions. (This is already true 
at NLO, i.e. $ F_2^H(x,Q^2) =0 \quad Q^2 < m_H^2, = \frac{\alpha_S}{4\pi}
C_{2,g}\otimes g^{n_f+1}(Q^2) Q^2>m_H^2$,
but the effect is very small.) This is 
a large effect at NNLO and is also negative at smallish $x$ 
($x  \sim 0.001$). Hence, ZMVFNS is not really feasible at NNLO, leading to
a huge discontinuity in $F^c_2(x,Q^2)$, which is significant in 
$F^{Tot}_2(x,Q^2)$, as shown in Fig. 1. 

\begin{figure}
  \includegraphics[height=.41\textheight]{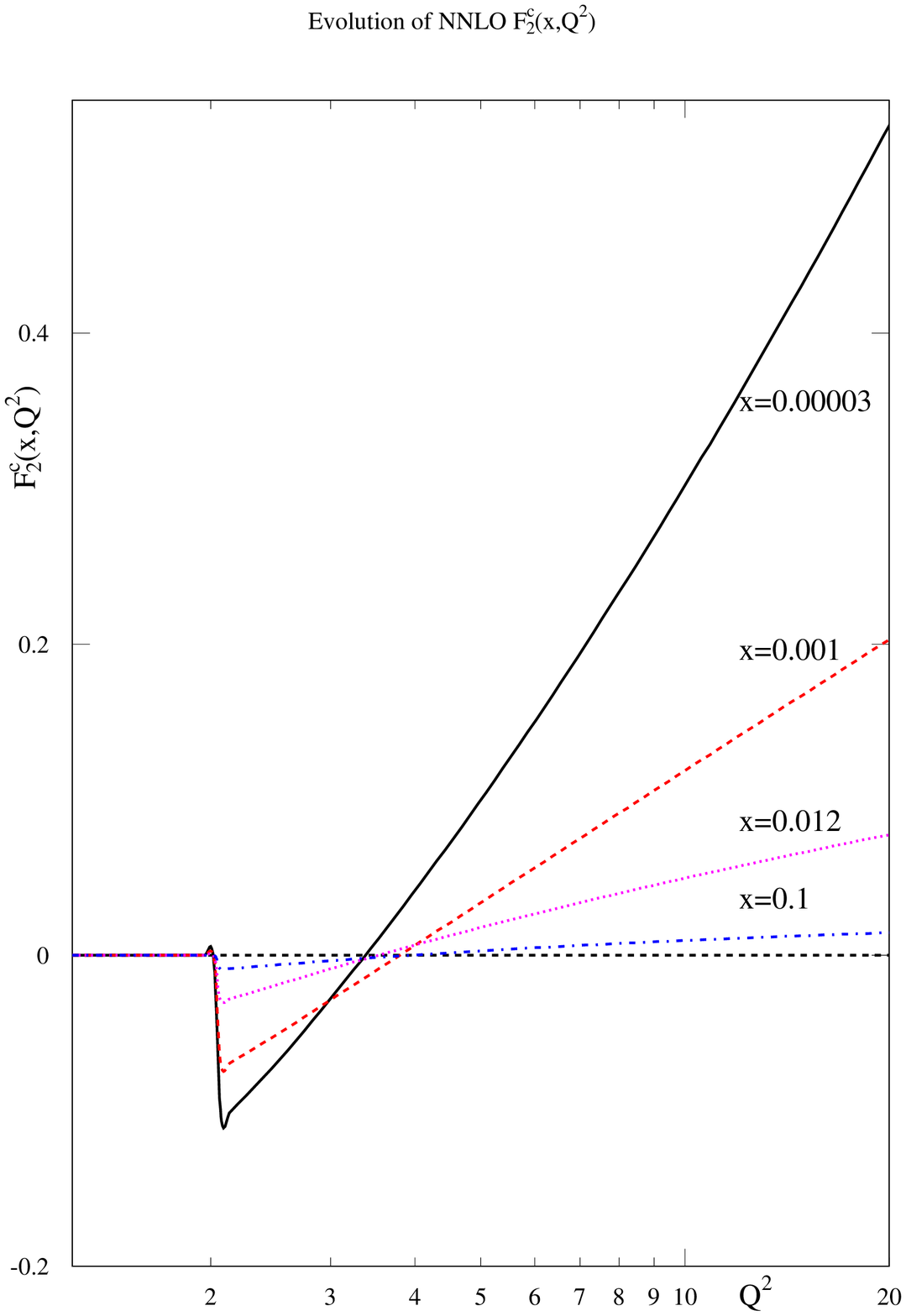}
\includegraphics[height=.41\textheight]{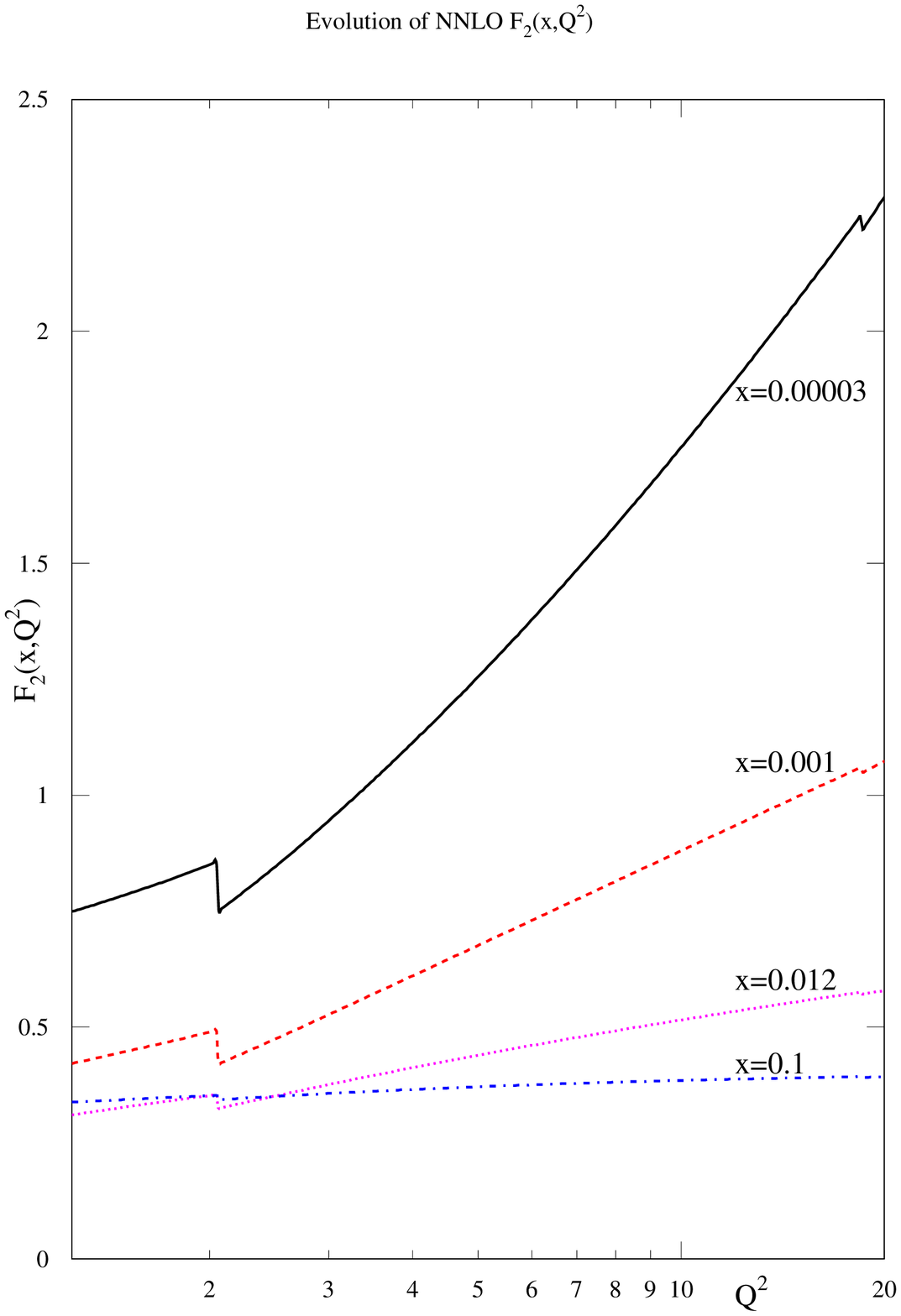}
\vspace{-5cm}
  \caption{NNLO $F_2^c(x,Q^2)$ and $F_2(x,Q^2)$ in zero-mass VFNS
\vspace{-0.7cm}}
\end{figure}

Hence we need a general Variable Flavour Number Scheme (VFNS)
interpolating between the two well-defined limits of $Q^2\leq m_H^2$ and 
$Q^2\gg m_H^2$. The VFNS can be defined by demanding equivalence of the 
$n_{f}$ and $n_f+1$-flavour descriptions at all 
orders,   
\begin{eqnarray}
F_i(x,Q^2)&=&C^{FF}_{i,k}(Q^2/m_H^2)\otimes f^{n_f}_k(Q^2)= 
C^{VF}_{i,j}(Q^2/m_H^2)\otimes f^{n_f+1}_j(Q^2)\nonumber \\
&\equiv& C^{VF}_{i,j}(Q^2/m_H^2)
\otimes A_{jk}(Q^2/m_H^2)\otimes f^{n_f}_k(Q^2)\nonumber \\
\to \quad C^{FF}_{i,k}(Q^2/m_H^2) &=& 
C^{VF}_{i,j}(Q^2/m_H^2)\otimes A_{jk}(Q^2/m_H^2).\nonumber
\end{eqnarray}
At ${\cal O}(\alpha_S)$ this gives
$$
C^{FF,1}_{2,g}(Q^2/m_H^2) = 
C^{VF,0}_{2, HH}(Q^2/m_H^2)\otimes P^0_{qg}\ln(Q^2/m_H^2)+
C^{VF,1}_{2,g}(Q^2/m_H^2).\nonumber
$$
The VFNS coefficient functions tend to the massless limits
as $Q^2/m_H^2 \to \infty$, as demonstrated to all orders in \cite{collins}, 
and if we use the zeroth order cross-section for 
photon-heavy quark scattering 
we obtain the original ACOT scheme \cite{acot}.

However, $C^{VF,0}_{2,HH}(Q^2/m_H^2)$ is only uniquely defined as  
$Q^2/m_H^2 \to \infty$, i.e. one can swap ${\cal O}(m_H^2/Q^2)$
terms between $C^{VF,0}_{2, HH}(Q^2/m_H^2)$ 
and $C^{VF,1}_{2,g}(Q^2/m_H^2)$. 
Similar reasoning holds for $C^{VF,n}_{2, HH}(Q^2/m_H^2)$. 
The ACOT prescription violated the threshold $W^2=Q^2(1-x)/x
> 4M^2$ since only one quark was needed in final state. The 
Thorne-Roberts variable flavour number scheme
(TR-VFNS) \cite{trvfns} recognized this ambiguity and removed it by 
imposing continuity of $(d\,F_2/d\,\ln Q^2)$ at the transition point.
This guaranteed smoothness at $Q^2=m_H^2$, but was complicated 
and cumbersome when extended to higher orders.  

There have been other alternatives, and most recently the 
ACOT($\chi$) prescription \cite{acotchi} defines  
$F^{H,0}_2(x,Q^2)=h^+(x/x_{max}, Q^2)$, where $x_{max}=
Q^2/(Q^2+4m_H^2)$. The coefficient functions tend to the massless 
limit for $Q^2/m_H^2 \to \infty$ but also respect the threshold requirement 
$W^2 \ge 4m_H^2$ for quark-antiquark production. Moreover it is 
very simple. For the VFNS to remain simple (and physical) at all 
orders I choose
$ C^{VF,n}_{2, HH}(Q^2/m_H^2,z)= C^{ZM,n}_{2, HH}(z/x_{max}).$\footnote{It is 
also important to choose
$C^{VF,n}_{L, HH}(Q^2/m_H^2,z)\propto C^{ZM,n}_{L, HH}(z/x_{max}).$} 
Adopting this convention then at NNLO we have, for example, 
$$
C^{VF,2}_{2, Hg}\biggl(\frac{Q^2}{m_H^2}\biggr)\!= C^{FF,2}_{2, Hg}\biggl(
\frac{Q^2}{m_H^2}\biggr) \!- 
C^{ZM,1}_{2, HH}\biggl(\frac{z}{x_{max}}\biggr)\otimes
A^1_{Hg}\biggl(\frac{Q^2}{m_H^2}\biggr)
\! -C^{ZM,0}_{2, HH}\biggl(\frac{z}{x_{max}}\biggr)\otimes
A^2_{Hg}\biggl(\frac{Q^2}{m_H^2}\biggr).\nonumber
$$
Since $A^2_{Hg}(1,z)\not=0$, $C^{2}_{2, Hg}(Q^2/m_H^2,z)$
is discontinuous at $Q^2=m_H^2$, and this compensates exactly for 
the discontinuity in the heavy flavour parton distribution.\footnote{At NNLO 
there are also contributions due to heavy flavours 
in loops away from the photon vertex. These are included within the VFNS and 
lead to a discontinuity in the coefficient functions
for light flavours  cancelling that in the light quark distributions. 
Strictly, part of this contribution should be interpreted as light flavour 
structure functions, while part of it 
contributes to $F^H_2(x,Q^2)$ \cite{smith}.} 

There is one more issue in defining the VFNS: the ordering for 
$F_2^H(x,Q^2)$, i.e.

\noindent \hspace{3cm} $n_f$-flavour \hspace{5cm} $n_f+1$-flavour

\vspace{0.2cm}

\noindent LO \hspace{2cm}$\frac{\alpha_S}{4\pi} 
C^{FF,1}_{2, Hg}\otimes g^{n_f}$\hspace{5cm}
$C^{VF,0}_{2, HH}\otimes h^+$

\noindent NLO \hspace{1cm}$\biggl(\frac{\alpha_S}{4\pi}\biggr)^2
(C^{FF,2}_{2, Hg}\otimes g^{n_f}+C^{FF,2}_{2, Hq}\otimes \Sigma^{n_f})$
\hspace{1cm}$\frac{\alpha_S}{4\pi}(C^{VF,1}_{2, HH}\otimes h^+
+C^{FF,1}_{2, Hg}\otimes g^{n_{f+1}}).$

\vspace{0.2cm}

\noindent Switching directly when going from 
$n_f$ to $n_f+1$ flavours leads to a discontinuity.
We must decide how to deal with this.
Up to now ACOT have used e.g. 
at NLO 
$$
\frac{\alpha_S}{4\pi} 
C^{FF,1}_{2, Hg}\otimes g^{n_f} \to \frac{\alpha_S}{4\pi}
(C^{VF,1}_{2, HH}\otimes h^+
+C^{FF,1}_{2, Hg}\otimes g^{n_f+1})+C^{VF,0}_{2, HH}\otimes h^+,\nonumber
$$
i.e. the same order of $\alpha_S$ above and below,
but LO below and NLO above. The Thorne-Roberts scheme proposed e.g. at LO 
$$
\frac{\alpha_S(Q^2)}{4\pi} 
C^{FF,1}_{2, Hg}\biggl(\frac{Q^2}{m_H^2}\biggr)\otimes g^{n_f}(Q^2) 
\to \frac{\alpha_S(m_H^2)}{4\pi} 
C^{FF,1}_{2, Hg}(1)\otimes 
g^{n_f}(m_H^2)+ C^{VF,0}_{2, HH}\biggl(\frac{Q^2}{m_H^2}\biggr)\otimes 
h^+(Q^2)\nonumber
$$
i.e. the higher order $\alpha_S$ term is frozen when going upwards 
through $Q^2=m_H^2$.
This difference in choice is extremely important at low $Q^2$.

\begin{figure}
  \includegraphics[height=.235\textheight]{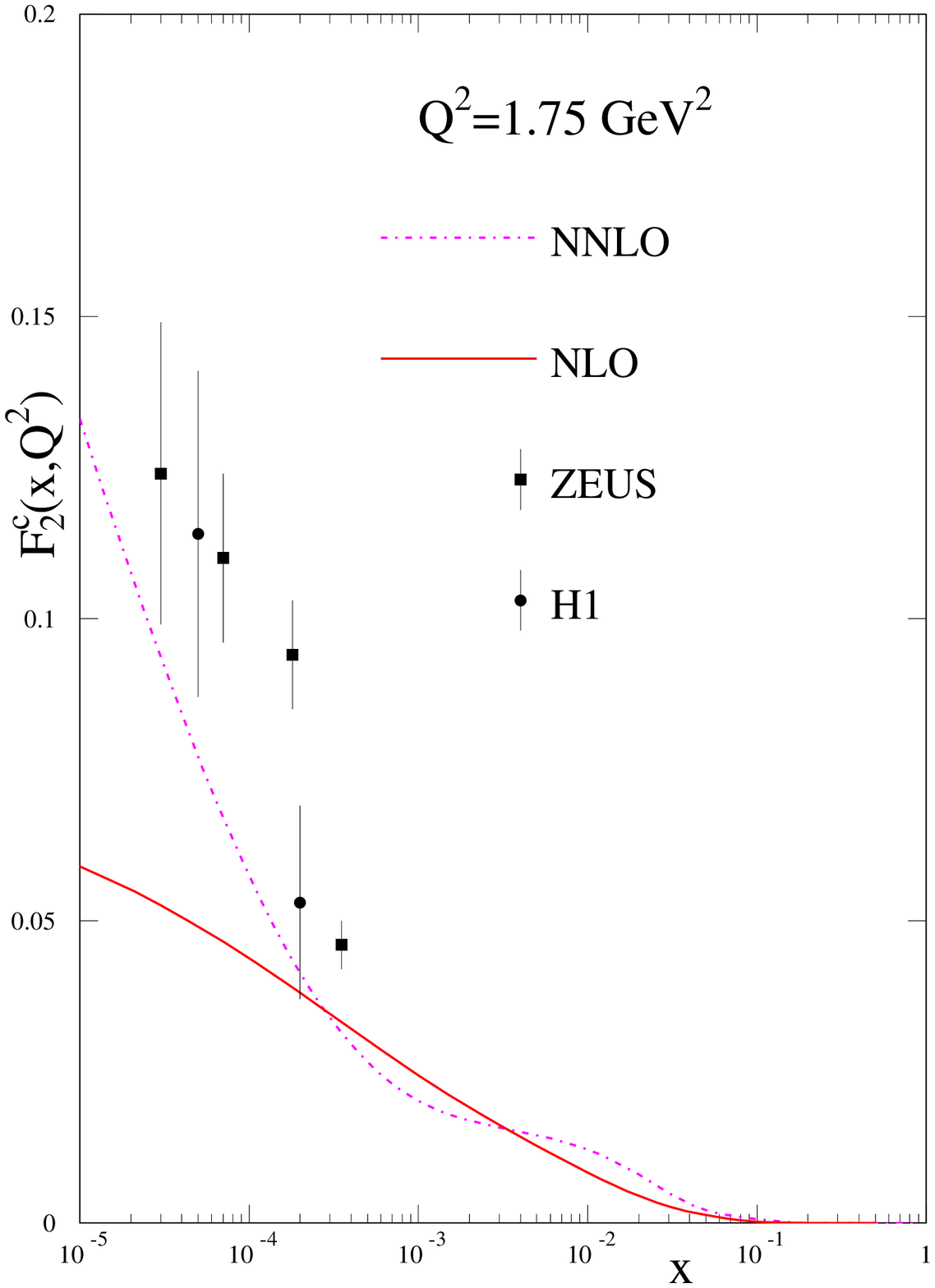}
\includegraphics[height=.235\textheight]{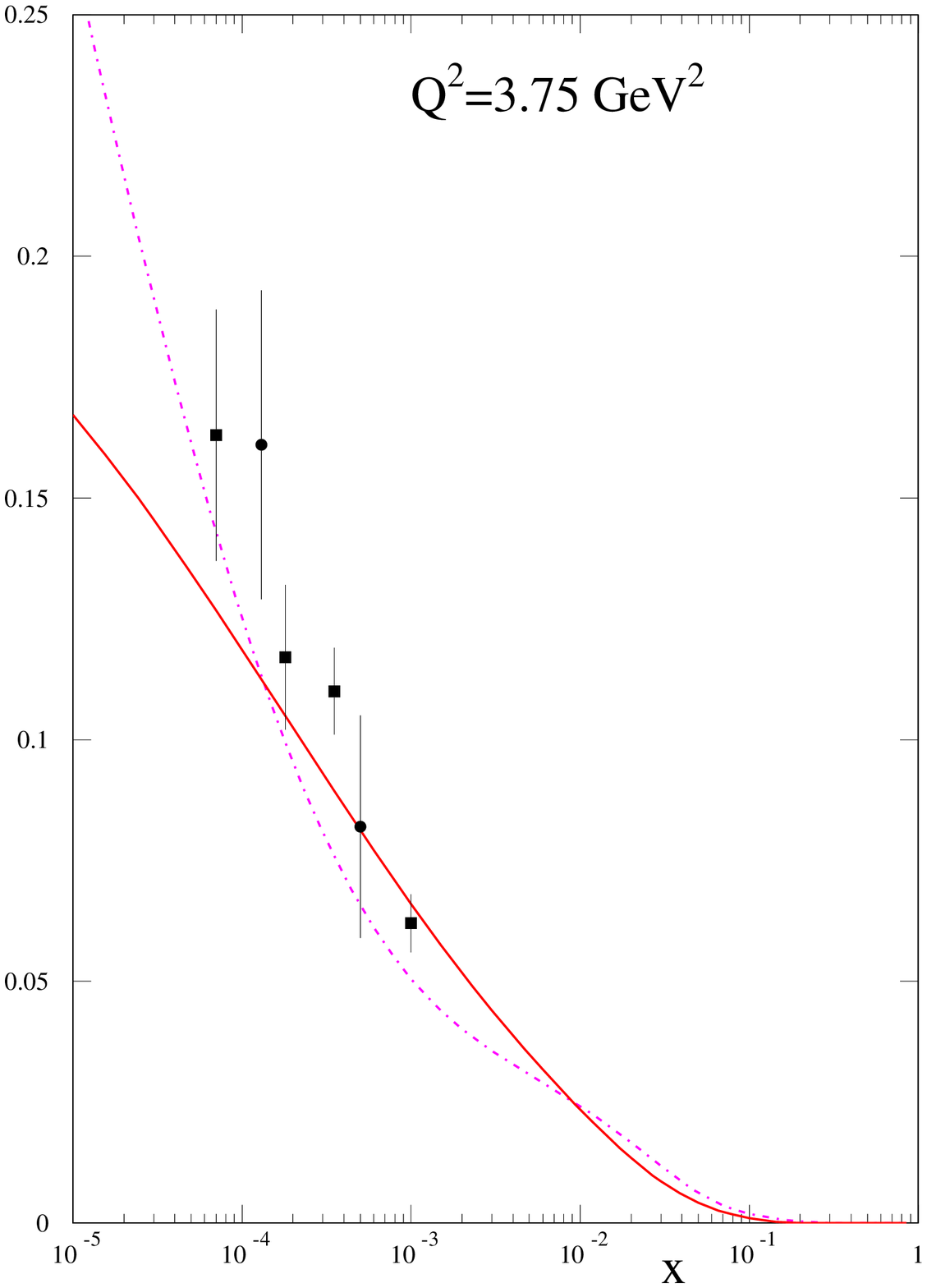}
  \includegraphics[height=.235\textheight]{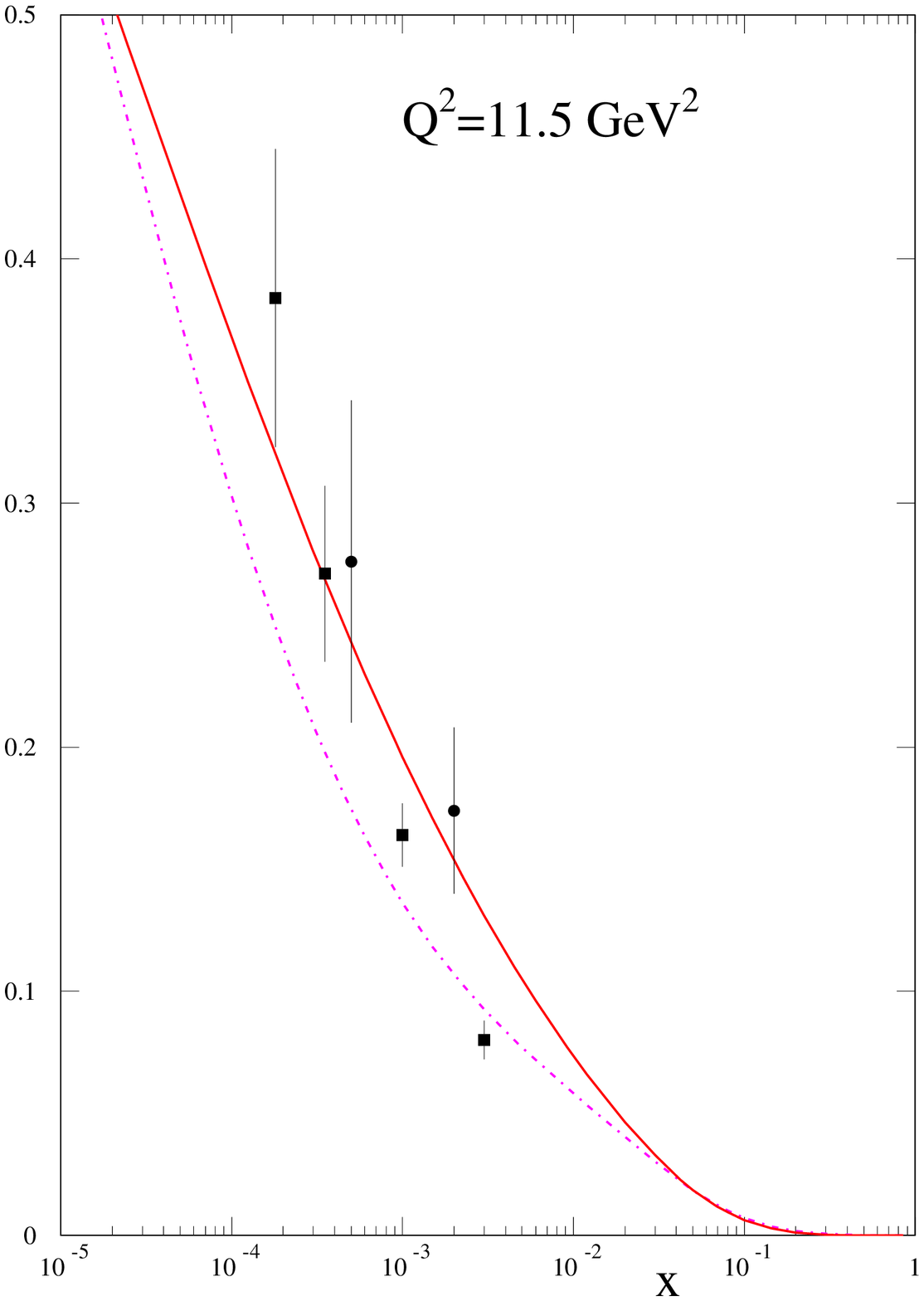}
\includegraphics[height=.235\textheight]{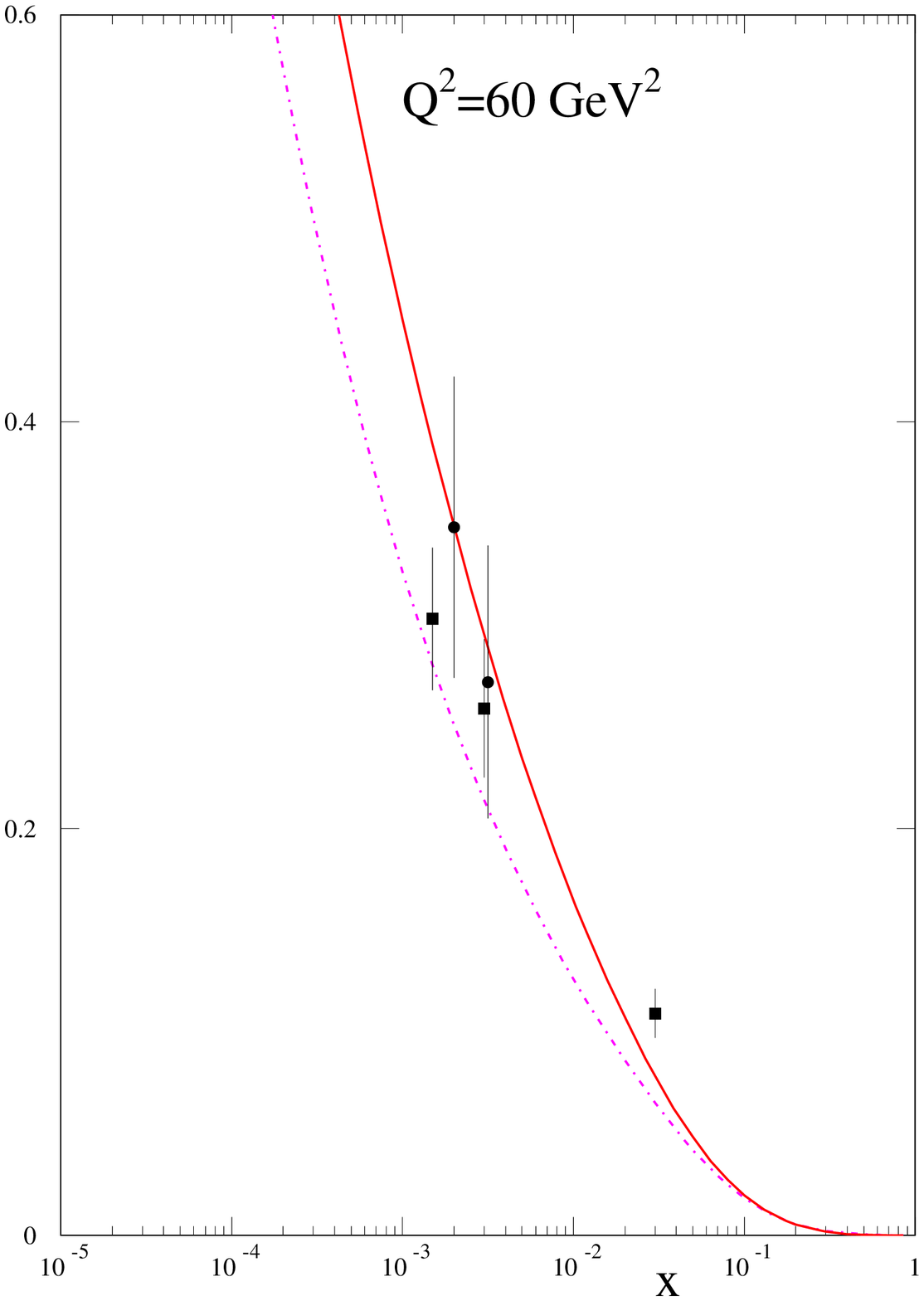}
  \caption{Comparison of NLO and NNLO predictions for $F_2^c(x,Q^2)$
\vspace{-0.7cm}}
\end{figure}

Making this choice, in order to define the VFNS at NNLO we need 
the ${\cal O}(\alpha_S^3)$ heavy flavour coefficient functions for 
$Q^2 \leq m_H^2$. However, these 
are not yet calculated (making a NNLO FFNS problematic). 
We know the leading threshold logarithms \cite{thresh},
and can derive the leading $ln(1/x)$ term from $k_T$-dependent
impact factors \cite{asymp},   
$$
C^{FF,3, low x}_{2, Hg}(Q^2/m_H^2,z) = 96\frac{\ln(1/z)}{z}
f(Q^2/m_H^2), \qquad f(1) \approx 4,\nonumber
$$
and $C^{FF,3, low x}_{2, Hq}(Q^2/m_H^2,z)=
4/9\,\,C^{FF,3, low x}_{2, Hg}(Q^2/m_H^2,z).$ 
By analogy with the known NNLO coefficient functions 
and splitting functions I hypothesize that
$$
C^{FF,3, low x}_{2, Hg}(Q^2/m_H^2,z) = \frac{96}{z}(\ln(1/z)-4)
(1-z/x_{max})^{20}f(Q^2/m_H^2),\nonumber
$$
i.e. the leading $\ln(1/z)$ term is always accompanied by $\sim -4$, 
and the effect of the
small $z$ term is damped as $z \to 1$.  Using the full 
(if slightly approximate) VFNS one 
can produce NNLO predictions for charm with 
discontinuous partons, but a continuous $F^c(x,Q^2)$. 
NNLO clearly improves the match to lowest $Q^2$ data \cite{zeus,h1}, where NLO 
is generally too low, as seen in Fig. 2.  

Hence, I have devised a full NNLO VFNS, with  a small amount 
of necessary  modelling. This seems to improve the fit to the lowest $x$ 
and $Q^2$ data greatly.
It also guarantees continuity of the physical observables, such as structure 
functions, despite the discontinuity in NNLO parton distributions. 
It can now be used in a full NNLO global analysis.


\end{document}